\documentclass[%
floatfix,
 aip,
amsmath,amssymb,
reprint,%
]{revtex4-1}

\usepackage{graphicx}
\usepackage{dcolumn}
\usepackage{bm}
\usepackage{multirow}

\usepackage[utf8]{inputenc}
\usepackage[T1]{fontenc}
\usepackage{mathptmx}
\usepackage{etoolbox}
\usepackage{physics}
\usepackage{siunitx}
\usepackage{mathtools}
\usepackage{titlesec}
\usepackage{upgreek}
\usepackage{xr}

\makeatletter
\def\@email#1#2{%
 \endgroup
 \patchcmd{\titleblock@produce}
 {\frontmatter@RRAPformat}
 {\frontmatter@RRAPformat{\produce@RRAP{*#1\href{mailto:#2}{#2}}}\frontmatter@RRAPformat}
 {}{}
}%
\makeatother
\begin{document}
\titlespacing\section{5pt}{8pt plus 2pt minus 2pt}{5pt plus 2pt minus 2pt}
\titlespacing\subsection{5pt}{8pt plus 2pt minus 2pt}{5pt plus 2pt minus 2pt}
\titlespacing\subsubsection{5pt}{8pt plus 2pt minus 2pt}{5pt plus 2pt minus 2pt}

\preprint{AIP/123-QED}

\title{Partial Auger Decay Widths from Complex-Valued Density Matrices}
\author{Florian Matz}
\author{Angelos Gkogkos}
\author{Thomas-C. Jagau}%
\email{florian.matz@kuleuven.be}
\email{thomas.jagau@kuleuven.be}
\affiliation{Department of Chemistry, KU Leuven, B-3001 Leuven, Belgium}
\DeclareSIUnit\angstrom{\text {Å}}

\date{\today}

\begin{abstract}
We discuss a new strategy to compute partial Auger decay widths with equation-of-motion ionisation-potential coupled-cluster (EOMIP-CCSD) wave functions in the framework of non-Hermitian quantum mechanics, where the decaying character of the metastable states is described via complex-scaled basis functions. The EOMIP-CCSD approach is universally applicable to many different electronic states and molecules that are Feshbach resonances, as proven by a multitude of previous studies. However, while the total decay width can generally be obtained from the energy eigenvalues, the computational of partial decay widths, i.\,e. the contributions of different channels to the total decay rate, which governs the probability distribution of the formation of different states in the decay process, is less trivial. In the past, methods where channels are projected out during the EOMIP-CCSD iteration have been developed (Auger Channel Projectors), but such a procedure requires to establish convergence of the excitation vector for each excitation manifold separately. Furthermore, they suffer from interaction between the channels upon perturbation of the wave function.

In contrast, we suggest to compute the contribution of the two-electron transition that Auger decay implies, where two valence electrons are involved, one refilling the core hole and one emitted to the continuum, from the respective entries in the two-electron density matrix that describe the extent of this transition in the wave function upon application of correlation and excitation operator. In this way, we obtain all partial decay widths from wave functions determined in the full excitation manifold. The results from this approach compare very well to the Auger Channel Projector results: we compute spectra for K-edge ionised states of methane, ethane, hydrogen sulfide, and the cyanide anion, as well as Coster-Kronig spectra of L$_1$-edge ionised hydrogen sulfide, which differ only negligibly between the two methods. A spectrum of the cyanide anion has not been reported before -- we discuss the selectivity of the decay process with respect to the initial state and the possibility of interatomic Auger decay.
\end{abstract}


\maketitle

\section{Introduction}

The equation-of-motion ionisation-potential coupled-cluster (EOMIP-CCSD) method~\cite{stanton94} is a highly versatile method for the description of ionised states. It combines an accurate treatment of correlation through the coupled cluster excitation operator with the possibility to solve for several states at once and to compute transition properties between different states or ionised states and the ground state.

Recently, it has been shown that in the framework of non-Hermitian quantum chemistry, EOMIP-CCSD calculations with complex-scaled basis sets are capable of reproducing the decay of the electronic molecular Feshbach resonances that are produced when ionising electrons from deeper levels.~\cite{matz22a,matz23a,parravicini23,drennhaus24} These states are resonances because they can decay via deexcitation of a valence electron into the vacancy, causing ionisation of a second electron with the excess energy. This process is called Auger decay~\cite{auger23} when it happens within one molecule or atom, and interatomic/intermolecular Coulombic decay~\cite{cederbaum97,marburger03} when it happens in a cluster of multiple atoms or molecules, where the deexcitation and the ionisation happen in different subsystems. All of these processes are sources of electrons with relatively low energy that can interact with the environment in various ways, which happen to some extent under any X-ray irradiation and after certain nuclear decay processes.

Non-Hermitian quantum mechanics itself~\cite{moiseyev2011non} is a term for methods that try to compute complex energy eigenvalues, which describe the time-dependent behaviour of the state besides its energy. In the family of methods we are applying, this is realised by using complex-scaled exponents in the basis functions.~\cite{rescigno78,moiseyev79} Besides computing the total decay width and energy of the ionised states, a particular interest lies in the computation of the partial widths, which determine the ratios between different final states the system can decay to depending on the electrons involved in the decay process. Due to energy conservation, this also determines the distribution of the energy of the outgoing electrons, which can be recorded in the form of Auger spectra.~\cite{mehlhorn98}

For the computation of partial widths, we have previously implemented and benchmarked a projected EOMIP-CCSD method.~\cite{matz23a} In this procedure, which we refer to as Auger Channel Projection (ACP), particular excitations describing Auger decay, i.\,e. those without a core-hole, but with two vacancies in the valence shell and one electron in a virtual orbital, are projected out from the wave function. This approach is related to core-valence projection~\cite{cederbaum80}, where all excitations that leave the system without a core-hole are projected out, in order to represent a core-vacant state in a bound fashion.~\cite{coriani15} However, the ACP are channel-specific and typically applied to project out only one channel at a time. With these projections, the state is still unbound, but the total decay width is reduced due to the absence of one channel. The difference in total width is approximating the partial width of the channel that has been projected out.

This approach makes it necessary to run multiple calculations for each molecule, one for each possible decay channel. Depending on the number of necessary iterations, this can lead to large computational costs. This is even more relevant for systems with multiple core orbitals, which each lead to a distinct set of partial decay widths. Examples for such systems include firstly molecules with multiple identical atoms from the second period or higher, which have core-holes very close in energy. These can be delocalised over the atoms due to symmetry. Typically, such core orbitals can not selectively be excited.

Secondly, systems with different atoms from the second period or higher have multiple core orbitals with relatively large energetic separation. For example, the core-ionisation energy of oxygen atoms is ca. 520\,eV, of nitrogen atoms ca. 400\,eV and of carbon atoms ca. 285\,eV. These states can selectively be produced by ionising radiation with the appropriate energy. Holes in lower-lying edges of such molecules can also decay via interatomic Auger decay, which are transitions between core electrons of different atoms. These have been measured both in solid state~\cite{gallon70} and in molecules~\cite{ogurtsov07}, but typically have very low intensity.

A third kind of molecules with multiple core-holes are those that contain elements of the third row of the periodic table or higher. In these, not only the K-shell, but also the L-shell or higher energy levels are sufficiently stabilised to initiate intramolecular decay when they are ionised. This includes the very fast Coster-Kronig decay~\cite{coster35}, where one of the electrons involved in the decay process stems from the same shell as the initial hole, for example in a L$_\mathrm{1}$ hole that is refilled from the L$_\mathrm{2,3}$ subshell.

Auger decay in all these systems has been studied with complex-scaled basis functions and ACP-EOMIP-CCSD before~\cite{matz22a,matz23a,matz23b,drennhaus24}, but such calculations are often computationally demanding. The motivation of this work is to develop a more favorable approach for the computation of partial widths for EOMIP-CCSD states, in which there is no need to repeat the calculation for every channel. Instead, we aim to solve for the complete wave function including contributions from all channels and extract partial widths from the contributions to this wave function similar to the energy decomposition approach for CCSD wave functions with core-holes, where widths are extracted from the contribution of double excitations to the energy. A previous attempt for such a decomposition has been made based on recomputation of the density matrix, but failed with nonphysical results due to the dependency of the density matrix elements on products among excitation operators. In this work, we want to revisit this approach and show how the blockwise analysis of the density matrix allows the extraction of physically sound partial widths.

\section{Theory}

EOM-CC wave functions are constructed by applying a linear excitation operator to the coupled cluster~\cite{cizek66,cizek69,cizek71} wave function:~\cite{ghosh84,sekino84,koch90,kaldor91,stanton93}

\begin{equation}\label{eq:eom-wf}
        |\Psi_\mathrm{EOMCC}\rangle = \hat{R}\mathrm{e}^{\hat{T}}|\Psi_0\rangle
\end{equation}

Solving the Schrödinger equation for a wave function parametrised like this corresponds to solving the Schrödinger equation with a similarity-transformed Hamiltonian $\bar{H} = \mathrm{e}^{-\hat{T}}\hat{H}\mathrm{e}^{\hat{T}}$:

\begin{equation}\label{eq:eom-sgl}
        \bar{H}\hat{R}|\Psi_0\rangle = E\hat{R}|\Psi_0\rangle
\end{equation}

This Hamiltonian is non-Hermitian, which implies that its eigenvalues can be complex and its left and right eigenvectors are not simply complex conjugates. To each eigenvalue $E$, there is a left eigenvalue equation

\begin{equation}
        \langle\Psi_0|\hat{L}^\dagger\bar{H} = \langle\Psi_0|\hat{L}^\dagger E
\end{equation}

with a distinct solution $\langle\Psi_\mathrm{EOMCC}| = \langle\Psi_0|\hat{L}^\dagger\mathrm{e}^{-\hat{T}}$. The wave functions are biorthogonal to each other: their normalisation condition is $\langle\Psi_\mathrm{EOMCC}|\Psi_\mathrm{EOMCC}\rangle = \langle\Psi_0|\hat{L}^\dagger\hat{R}|\Psi_0\rangle = 1$.

In the EOM-CCSD method, which is used in this work, the excitation operators are truncated to the first two terms. This implies that $\hat{R} = \hat{R}_1 + \hat{R}_2$, $\hat{T} = \hat{T}_1 + \hat{T}_2$, and $\hat{L} = \hat{L}_1 + \hat{L}_2$.  In the EOMIP-CCSD method, $\hat{R}$ is chosen such that it reduces the number of electrons in the system by one.~\cite{stanton94} It is thus defined as

\begin{equation}
        \hat{R}_\mathrm{IP} = \sum_i^{N_\mathrm{e}} r_i \hat{i} + \frac{1}{2} \sum_{ij}^{N_\mathrm{e}} \sum_a^{N_\mathrm{v}} r_{ij}^a \hat{a}^\dagger \hat{i} \hat{j}.
\end{equation}

In this manuscript, the operator $\hat{L}$ for the left-side wave function is written as another excitation operator

\begin{equation}
        \hat{L}_\mathrm{IP} = \sum_i^{N_\mathrm{e}} l_i \hat{i} + \frac{1}{2} \sum_{ij}^{N_\mathrm{e}} \sum_a^{N_\mathrm{v}} l_{ij}^a \hat{a}^\dagger \hat{i} \hat{j}
\end{equation}

but becomes a deexcitation operator or an excitation operator acting towards the left upon complex conjugation.

For simply computing energies, it is sufficient to solve for one of the two wave functions, since both energy eigenvalues are identical. However, some properties such as energy derivatives can only be evaluated as expectation values which requires the left and the right wave function. Quantities that result from the coupling of different states, such as spin-orbit coupling, are evaluated as transition moments: integrals over a left wave function, an operator and a right wave function.

A convenient and generalisable way to express such expectation values and transition moments is via density matrices. They can be understood as the contributions of excitations of a specific rank to a wave function. For example, the two-particle density matrix, which is defined as

\begin{equation}\label{eq:dm}
        \rho_{pq}^{rs} = \langle\Psi_0|\hat{L}^\dagger\mathrm{e}^{-\hat{T}}\{\hat{p}^\dagger\hat{q}^\dagger\hat{r}\hat{s}\}\mathrm{e}^{\hat{T}}\hat{R}|\Psi_0\rangle,
\end{equation}

measures the two-particle excitation character of the correlated wave function relative to the Hartree-Fock ground state. In this equation, $p$, $q$, $r$, and $s$ are any of the $N_\mathrm{orb}$ molecular orbitals, either occupied or unoccupied. 
Any expectation value can be expressed as a product of matrix elements of the operator between substituted determinants, and these density matrices. It is notably to mention that for most quantities, only certain ranks of density matrices contribute depending on the character of the operator. The Hamiltonian, for example, only has one- and two-electron parts, such that the energy difference between Hartree-Fock state and reference state can be written as a function of one- and two-particle density matrices as

\begin{widetext}\begin{equation}
        E-E_\mathrm{HF} = \langle\Psi|\hat{H}|\Psi\rangle = \sum_{pq}^{N_\mathrm{orb}} \langle \Psi|\hat{p} f_{pq} \hat{q}^\dagger |\Psi\rangle + \frac{1}{4}\sum_{pqrs}^{N_\mathrm{orb}} \langle\Psi|\hat{p}\hat{q}\langle pq||rs\rangle \hat{r}^\dagger\hat{s}^\dagger|\Psi\rangle
         = \sum_{pq}^{N_\mathrm{orb}} f_{pq}\rho_p^q + \frac{1}{4} \sum_{pqrs}^{N_\mathrm{orb}} \rho_{pq}^{rs} \langle pq||rs\rangle
\end{equation}\end{widetext}

with the Fock matrix elements $f_{pq}$ and the two-electron integrals $\langle pq||rs\rangle$.

Our goal is to use this density matrix representation of the energy to compute partial Auger decay widths. In general, the decay of electronic resonances can be described by non-Hermitian quantum mechanics~\cite{moiseyev2011non,jagau22}, where complex energy eigenvalues~\cite{siegert39}

\begin{equation}
        E_\mathrm{res} = E_\mathrm{R} - \mathrm{i}\frac{\Gamma}{2}
\end{equation}

are computed. These represent not only the energy $E_\mathrm{R}$, but also the decay width $\Gamma$ of resonances. Such complex energies can be obtained by implicitly taking the outgoing boundary conditions into account, as has been done by using complex-scaled Gaussian basis functions (CBF)~\cite{rescigno78,mccurdy78}, with exponents that have been multiplied by a complex number $\mathrm{e}^{-2\mathrm{i}\theta}$.

The imaginary part of Hartree-Fock energies of bound states is negligible and an artifact present only with a finite basis set, but core-excited states that include determinants describing the decay of the state lead to substantial imaginary contributions to the energy.~\cite{matz22a} For example, the operator $r_{ij}^{a} \hat{a}^\dagger \hat{i} \hat{j}$, when applied to the reference state, can create two holes in valence orbitals and one particle in the virtual space. Such contributions are included when carrying out EOMIP-CCSD calculations that produce core-ionised states. The two holes in valence orbitals represent the possible electronic configurations after Auger decay, and the particle in the virtual space represents the outgoing Auger electron.

To compute partial widths, one can project these wave functions excitation operators on manifolds excluding determinants that describe certain decay channels and evaluate their energy without one channel present. The difference in energy is then taken as partial width of that channel.

One possibility for this is the recomputation of density matrices with these projected excitation operators to evaluate the energy of the state.~\cite{matz22a} It was shown that this approach is successful for CCSD wave functions, but leads to incorrect partial widths for EOMIP-CCSD states. This could be explained by the products of $\hat{R}$, $\hat{L}$, and $\hat{T}$ that appear in equation~\ref{eq:dm}: the parts of $\hat{R}$ and $\hat{L}$ describing Auger transitions also contribute to other determinants and $\hat{T}$ can also create determinants in the wave function that represent Auger decay, even though no part $\hat{T}$ can uniquely be assigned to each decay channel. Through these products, much larger parts of the density matrix are perturbed as desired and the difference in energy does not correspond to partial widths.

Instead of this procedure, we propose to evaluate the effect of each channel on the decay width directly from the density matrix elements that describe the electronic transition that is happening during decay via that channel, namely the removal of two valence electrons from orbitals $i$ and $j$, and the creation of one electron in the core-hole $\textbf{c}$ and one electron in orbital $a$ of the virtual space. In this way, we circumvent the need to manipulate excitation operators. The relevant energy contribution from the density matrix is

\begin{align}\nonumber
        E_{ij}^{\textbf{c}a} &= \frac{1}{4} \langle ij||\textbf{c}a\rangle \langle\Psi_0|\hat{L}^\dagger \mathrm{e}^{-\hat{T}} \{\hat{i}^\dagger \hat{j}^\dagger \hat{\textbf{c}} \hat{a}\} \mathrm{e}^{\hat{T}} \hat{R}|\Psi_0\rangle\\
        &= \frac{1}{4} \langle ij||\textbf{c}a\rangle \rho_{ij}^{\textbf{c}a}
\end{align}

where the parts of the wave function that have holes in orbitals $i$ and $j$ ($\langle\Psi_\mathrm{EOMCC}|\hat{i}^\dagger\hat{j}^\dagger$) are projected onto the parts $\hat{\textbf{c}}\hat{a}|\Psi_\mathrm{EOMCC}\rangle$ of the wave function that have a hole in the core orbital $\textbf{c}$ and an unoccupied virtual orbital $a$. Later, the condition for the virtual space will vanish by summing over all possible $a$, since any orbital from the virtual space can describe the outgoing electron. This expression is comparable to the CCSD energy expression, where that energy contribution simplifies to

\begin{equation}
        E_{ij}^{\textbf{c}a} = \langle ij||\textbf{c}a\rangle \left(\frac{1}{4} t_{ij}^{\textbf{c}a} + \frac{1}{2} (t_i^\textbf{c} t_j^a + t_i^a t_j^\textbf{c})\right).
\end{equation}

The respective EOMIP-CCSD density matrix block has a more complex dependence on the excitation amplitudes: it is a sum of different products of amplitudes in $\hat{T}$, $\hat{L}$, and $\hat{R}$.~\cite{stanton94}

By summing over $a$, single partial widths are obtained as

\begin{equation}\label{eq:dmpw}
        -\frac{\Gamma_{ij}(\textbf{c})}{2} = \mathrm{Im}\left(\frac{1}{4} \sum_a^{N_\mathrm{v}} \langle ij||\textbf{c}a\rangle \rho_{ij}^{\textbf{c}a}\right)
\end{equation}

that only depends on the density matrix, and thus only on the excitation amplitudes.

In this work, results from equation~\ref{eq:dmpw} are compared against the only currently established approach to compute partial widths from EOMIP-CCSD wave functions for core-ionised states: the application of Auger Channel Projectors (ACP).~\cite{matz23a} In this method, determinants describing Auger decay are not projected out after convergence, but during the solution of equation~\ref{eq:eom-sgl}. The wave functions still depends on products of $\hat{R}$ and $\hat{T}$, but the perturbation of the wave function beyond the removal of the Auger decay channel can be balanced out because the other excitation amplitudes are optimised in the projected excitation manifold.

The resulting wave functions contain all Auger decay channels except for the one that has been projected out, which all contribute to an imaginary part that is typically slightly smaller than in the full excitation manifold. This difference can be assigned to the partial width of the projected channel. However, computing a partial decay width for one channel requires a complete and distinct solution of equation~\ref{eq:eom-sgl}, increasing the computational cost when one wants to compute all possible decay channels, the number of which scales with the square of the number of occupied orbitals. To extract the energy contributions from the density matrix as discussed above, only one right and one left wave function, each containing the full set of decay channels, need to be solved for. This also avoids the distortion of the computed partial widths due to the relaxation of the remaining amplitudes, which can potentially also involve changes in the partial widths of other decay channels.~\cite{matz23a}

In this work, we also employ the EOMDIP-CCSD method~\cite{nooijen97,sattelmeyer03} for the computation of doubly ionised states, which represent the final states of Auger decay. In this method, the excitation operator $\hat{R}$ in equation~\ref{eq:eom-wf} is defined as

\begin{equation}
        \hat{R}_\mathrm{DIP} = \frac{1}{2} \sum_{ij}^{N_\mathrm{e}} r_{ij} \hat{i}\hat{j} + \frac{1}{6}\sum_{ijk}^{N_\mathrm{e}} \sum_a^{N_\mathrm{v}} r_{ijk}^{a} \hat{a}^\dagger\hat{i}\hat{j}\hat{k},
\end{equation}

such that it removes two electrons from the system. The double ionisation amplitudes $r_{ij}$ describe the contribution of certain determinants with two vacancies to the possible doubly ionised states.

\section{Computational Details}

The calculations were carried out with a modified version of the Q-Chem 6.1~\cite{epifanovsky21} program package. We used geometries and basis sets equivalent to those used in our earlier studies of methane~\cite{matz23b}, ethane~\cite{matz23b}, and hydrogen sulfide~\cite{drennhaus24}, as detailed in the following paragraphs. The exponents of the complex-scaled basis functions used in this work are also listed in the Supporting Information.

For the hydrocarbons, we used the ``cc-pCVTZ (5sp)'' basis set, including the s- and p-type basis functions from the cc-pCV5Z basis set and the d- and f-type functions from the cc-pCVTZ basis set. We augmented the basis set with 3 complex-scaled sets of s-, p-, and d-functions each, centered on every carbon atom. The exponents of two of these functions were optimised for the neon atom~\cite{matz22a} and scaled according to the different nuclear charge. A third complex-scaled function per angular momentum was added on each atom to take into account the more diffuse electron distribution, the exponent of which is half of the exponent of the more diffuse of the two optimised functions. Methane was calculated in the T$_\mathrm{d}$ point group with a bond length $d_\mathrm{CH}=1.0905 \si{\angstrom}$. Staggered ethane was calculated in the D$_{3\mathrm{d}}$ point group with bond lengths $d_\mathrm{CC}=1.5326 \si{\angstrom}$ and $d_\mathrm{CH}=1.0968 \si{\angstrom}$, and a bond angle $\alpha_\mathrm{CCH}=111.33^\circ$.

The calculations on H$_2$S were carried out with the aug-cc-pCVTZ (5sp) basis set, where 4 complex-scaled s-, p-, and d-shells centered on the sulfur atom were added to compute the decay of the K-edge vacancy. The exponents of these were chosen as the two optimised basis functions, one that is, such as in the hydrocarbons, by a factor of 2 more diffuse than the more diffuse optimised function, and one function with an exponent that is the geometric average of the two optimised exponents. For the calculations on the L$_\mathrm{1}$-edge vacancy, 8 complex-scaled s-, p-, and d-shells centered on the sulfur atom were used. This was proven necessary before to quantitatively capture the Coster-Kronig decay, which causes the emission of electrons in a very different energy range compared to normal Auger decay.~\cite{drennhaus24} The exponents of these functions were chosen as the four complex-scaled shells used for the K-edge vacancy computation, and four additional functions per angular momentum. The exponents for these were determined with an even-tempered spacing starting from the most diffuse of the four complex-scaled shells used before with a spacing factor of 2. H$_2$S was calculated in the C$_{2\mathrm{v}}$ point group with the bond length $d_\mathrm{SH} = 1.3338 \si{\angstrom}$ and the bond angle $\alpha=92.205^\circ$.

The decay of the cyanide anion was, similar to the hydrocarbons, computed with the cc-pCVTZ (5sp) basis set, with three shells of complex-scaled s-, p-, and d-functions each on both the carbon and the nitrogen atom. An interatomic distance of $d_\mathrm{CN} = 1.1241 \si{\angstrom}$ was used.

For all states, the partial widths were determined at the optimal scaling angle $\theta$. The optimal scaling angle is determined by minimising the dependence of the energy on the scaling angle $\left|\mathrm{d}E/\mathrm{d}\theta\right|$~\cite{jagau17,moiseyev78}, which is realised by computing $E(\theta)$ in steps of 1$^\circ$ in a range from 0$^\circ$ to 45$^\circ$. For methane, ethane, and hydrogen sulfide, these optimal angles were determined before. They are reported at the beginning of the discussion of the respective results.

The energies of the final states of Auger decay were computed with the EOMDIP-CCSD approach. The used basis sets were equivalent to the basis sets for the EOMIP-CCSD calculations, but without the complex-scaled basis functions. Earlier work~\cite{drennhaus24arxiv} showed that some final states of the decay of the K-edge vacancy in H$_2$S cannot be converged using the EOMDIP-CCSD method. This affects all states with a hole in the $L_\mathrm{1}$-edge orbital. We used the extrapolation procedure from that work, where the EOMDIP-CCSD energies are computed for all states with a reduced excitation manifold including only determinants with 2 holes, and for those not affected by the convergence problem with the full excitation manifold that also includes 3-hole-1-particle terms. We then used a linear extrapolation to predict the correlation energies of the determinants with L$_\mathrm{1}$-edge vacancies.

To map the results of the partial widths calculations, which are obtained for single pairs of Hartree-Fock valence orbitals, onto the results of the EOMDIP-CCSD calculation, the weighting procedure established in~\cite{matz23b} was used. For this, the partial widths for the two-hole states are multiplied with the corresponding squared amplitude $r_{ij}^2$ in the EOMDIP-CCSD excitation vector. Gaussian peaks with areas proportional to that number are centered at the energies computed for the corresponding EOMDIP-CCSD solution according to

\begin{equation}
        E_\mathrm{Auger} = E_\mathrm{IP}(\mathrm{A}^{+**}) - E_\mathrm{DIP}(\mathrm{A}^{2+}).
\end{equation}

For the 1a$_1$ core-hole in H$_2$S, this weighting procedure would lead to inconsistencies due to the presence of the extrapolated DIP energies. For these states that were converged without the presence of 3-hole-1-particle amplitudes, $r_{ij}^2$ for each state always add up to 1, while this is not the case for states that were computed in the larger excitation manifold. Because of this, these spectra was determined with $\displaystyle r_{ij}^2/\sum_{ij}r_{ij}^2$ as weighting factor.

\section{Results}

\subsection{Methane}

The optimal scaling angle for core-ionised methane with the EOMIP-CCSD wave function and the present basis set was determined before to be 18$^\circ$, with which the total decay half-width results as 40.1\,meV.~\cite{matz23b} The $\Delta$CCSD method computes the value as 43.4\,meV~\cite{matz23b}, comparable to the experimental value of $48 \pm 1$\,meV.~\cite{carroll99}

The partial widths computed with ACP-EOMIP-CCSD, from decomposing the CCSD energy and the present results from decomposing the density matrix are listed in table~\ref{tab:methanedmdecomp}.

\begin{table}\centering
        \caption{Partial half-widths for the Auger decay of the core-ionised methane molecule, computed with different methods.}\label{tab:methanedmdecomp}
\begin{tabular}{lrrr}\hline
        & \multicolumn{3}{c}{Partial widths / meV}\\
        Decay & ACP- & \multicolumn{2}{c}{Energy decomposition}\\
        channel & EOM~\cite{matz23b} & CCSD~\cite{matz23b} & EOMIP-CCSD\\\hline
        $^1$A$_1$ (2a$_1^{-2}$)&6.0&6.7&7.7\\
        $^1$T$_2$ (2a$_1^{-1}$1t$_2^{-1}$)&10.4&15.1&12.2\\
        $^3$T$_2$ (2a$_1^{-1}$1t$_2^{-1}$)&2.8&4.7&3.0\\
        $^1$A$_1$ (1t$_2^{-2}$)&0.9&1.6&0.9\\
        $^1$T$_2$ (1t$_2^{-2}$)&11.6&15.4&11.8\\
        $^1$E (1t$_2^{-2}$)&7.3&11.2&7.4\\
        $^3$T$_1$ (1t$_2^{-2}$)&0.0&0.0&0.0\\\hline
        Sum&39.0&54.7&43.0\\\hline
\end{tabular}
\end{table}

The decay of the molecule is dominated by double ionisation of the 1t$_2$ orbitals, which is confirmed both by CCSD and EOMIP-CCSD calculations and can be explained by the threefold degeneracy of this level compared to the inner-valence, nondegenerate 2a$_1$-orbital. As discussed before, the CCSD method estimates decay widths involving 1t$_2$ orbitals significantly larger than the ACP-EOMIP method.

The decomposition of this energy leads to numbers very similar to the ACP approach with deviations of at most 1.8~meV. The largest deviation occurs in the $^1$T$_2$ (2a$_1^{-1}$1t$_2^{-1}$) channel where the new method brings the results closer to the CCSD numbers. In general, decay from the inner-valence orbital is estimated as more intense from the energy decomposition.

Still, the reproduction of the numbers obtained with ACP from the density matrix alone confirms that its elements allow to obtain reliable values contributions of the different decay channels, without a need to run separate projected calculations. It also implies that the difference to the CCSD results is not a flaw of the ACP method, but that the EOMIP-CCSD wave functions contains different contributions to the partial widths.

Both sets of EOM partial widths also add up to a number close to the computed total width, while the numbers obtained via decomposition of the CCSD energy sum up to a notably larger total width than the total decay width. This points to other terms in the CCSD wave function which do not correspond to Auger decay but reduce the decay width, which cancels out parts of the decay width caused by the Auger decay channels.

At this point it should be made clear that the partial widths obtained via decomposition of the density matrix have an important fundamental property: partial widths of single channels add up to the partial widths determined for multiple channels at once. This is because there is no relaxation effect of the other channels when projecting out one channel like in the ACP method. This implies that the difference between total width and sum of partial widths must either be nonphysical contributions to the decay due to the incompleteness of the basis set or decay channels not represented by the oocv block of the two-electron density matrix.

\begin{figure}
        \includegraphics[width=\linewidth]{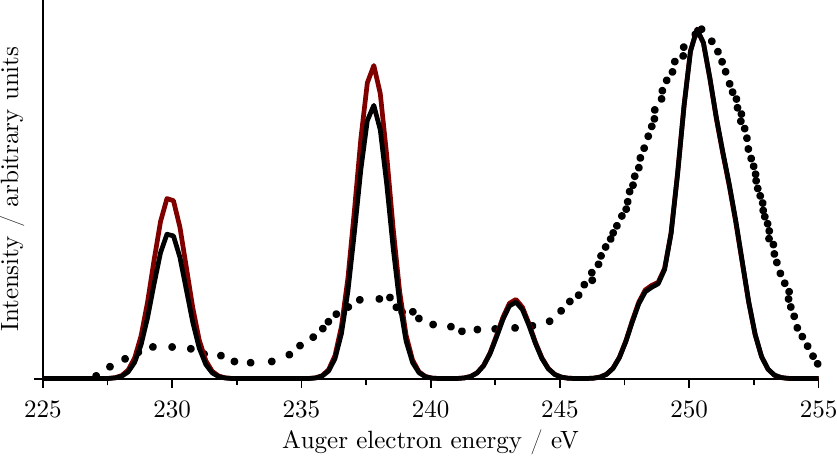}
        \caption{Auger decay spectrum of methane, computed with ACP-EOMIP-CCSD (black)~\cite{matz23b} and decomposition of the EOMIP energy (red), compared to an experimental spectrum~\cite{kivimaki96}. Full width at half maximum for the peaks is 1.5\,eV. Spectra shifted to lower energies by 1.2\,eV to match experiment.}\label{fig:methane-spek}
\end{figure}

With the double ionisation energies computed with EOMDIP-CCSD, we generated Auger spectra with the two sets of EOMIP-CCSD partial decay widths. These are shown in comparison to an experimental spectrum~\cite{kivimaki96} in figure~\ref{fig:methane-spek}.

The spectra reflect the large similarities of the two methods and the inner-valence channels being estimated as more intense by the energy decomposition method. A deviation to the experiment remains -- lower-energy signals are overestimated compared to the large peak at the high-energy end which arises from Auger decay involving two electrons from the 1t$_2$ HOMO. This is also not remedied by the weighting procedure, despite it slightly reducing the width of these low-energy peaks since 3-hole-1-particle terms have larger contributions to the corresponding EOMDIP-CCSD wave functions.

\subsection{Ethane}

The ethane molecule has two core-orbitals, one with an in-phase ($\textit{gerade}$) and one with an out-of-phase ($\textit{ungerade}$) combination of the carbon 1s orbitals. However, their ionisation energy only differs by 0.02\,eV, such that they cannot selectively be produced in an experiment. Their optimal $\theta$ values with EOMIP-CCSD are both 19$^\circ$.~\cite{matz23b} The total decay half-widths were calculated to 39.0 ($\textit{gerade}$) and 37.5 meV ($\textit{ungerade}$), and ca. 1\,meV lower with $\Delta$CCSD.~\cite{matz23b}

The partial widths were computed via energy decomposition. The sum of the partial widths in comparison with the total widths are given in table~\ref{tab:ethane-total} and the 12 most intense widths are listed in table~\ref{tab:ethanepw}.

\begin{table}\centering
\caption{Total Auger decay half-widths in meV for the two core holes of the ethane molecule, compared with the sum of partial decay widths, computed with different methods.}
\label{tab:ethane-total}
\begin{tabular}{llrrrr}\hline
        \multirow{3}{*}{Method}&\multicolumn{2}{c}{$\textit{gerade}$}&\multicolumn{2}{c}{$\textit{ungerade}$}\\
        &&$\Gamma$&$\Sigma\Gamma_i$&$\Gamma$&$\Sigma\Gamma_i$\\\hline
        \multirow{2}{*}{EOM-CCSD}&ACP~\cite{matz23b}&\multirow{2}{*}{39.0}&37.7&\multirow{2}{*}{37.5}&36.0\\
        &decomp.&&42.9&&40.6\\
        CCSD~\cite{matz23b}&&38.0&39.6&36.4&38.1\\\hline
\end{tabular}
\end{table}

A similar behaviour of the total widths occurs as in methane: the ACP method computes a too low sum of partial widths compared to the total EOMIP-CCSD decay width, while the energy decomposition method overestimates it, just like the decomposition of method does in the CCSD wave function. This implies that nonphysical contributions to the decay width, which are not captured in energy decomposition approaches, are typically reducing it, which is not captured in energy decomposition approaches. ACP partial decay widths are systematically too low because other decay channels can ``fill in'' for the missing channels, which increases their intensities, making the target channel apparently less intense.

\begin{table}\centering
\caption{Most intense partial decay half-widths of the two core holes of ethane, computed with different methods.}\label{tab:ethanepw}
\centering
\begin{tabular}{lrrrrrr}\hline
        & \multicolumn{6}{c}{Partial half-widths / meV}\\
        Decay &\multicolumn{2}{c}{ACP}&\multicolumn{4}{c}{energy decomposition}\\
        channel & \multicolumn{2}{c}{EOMIP-CCSD~\cite{matz23b}} & \multicolumn{2}{c}{CCSD~\cite{matz23b}} & \multicolumn{2}{c}{EOMIP-CCSD}\\
        & $\textit{g}$ & $\textit{u}$ & $\textit{g}$ & $\textit{u}$& $\textit{g}$ & $\textit{u}$\\\hline
        $^1$E$_u$ (1e$_u^{-1}$ 1e$_g^{-1}$) & 3.6 & 3.6 & 3.9 & 3.9 & 3.8 & 3.8\\
        $^1$E$_u$ (1e$_u^{-1}$ 3a$_{1g}^{-1}$) & 3.0 & 3.4 & 2.9 & 3.2 & 3.1 & 3.5\\
        $^1$E$_g$ (3a$_{1g}^{-1}$ 1e$_g^{-1}$) & 3.4 & 2.9 & 3.2 & 2.9 & 3.4 & 3.0\\
        $^1$A$_{2u}$ (2a$_{1g}^{-1}$ 2a$_{2u}^{-1}$) & 3.1 & 2.8 & 3.4 & 3.1 & 4.1 & 3.7\\
        $^1$A$_{1g}$ (3a$_{1g}^{-2}$) & 2.6 & 1.6 & 2.3 & 1.4 & 2.6 & 1.6\\
        $^1$E$_g$ (2a$_{1g}^{-1}$ 1e$_g^{-1}$) & 1.9 & 2.0 & 2.2 & 2.3 & 2.3 & 2.4\\
        $^1$E$_u$ (2a$_{1g}^{-1}$ 1e$_u^{-1}$) & 2.0 & 1.9 & 2.2 & 2.2 & 2.5 & 2.3\\
        $^1$A$_{1g}$ (2a$_{1g}^{-1}$ 3a$_{1g}^{-1}$) & 1.9 & 1.9 & 1.8 & 1.8 & 2.1 & 2.2\\
        $^1$E$_g$ (1e$_u^{-2})$ & 1.9 & 1.8 & 1.9 & 1.9 & 2.0 & 1.9\\
        $^1$A$_{1g}$ (2a$_{1g}^{-2}$) & 1.6 & 2.0 & 1.7 & 2.2 & 2.2 & 2.7\\
        $^1$E$_g$ (1e$_g^{-2}$) & 1.8 & 1.8 & 1.9 & 2.0 & 1.8 & 1.8\\
        $^1$A$_{2u}$ (2a$_{2u}^{-1}$ 3a$_{1g}^{-1}$) & 2.1 & 1.2 & 1.9 & 1.3 & 2.3 & 1.4\\\hline
\end{tabular}
\end{table}

Most of the partial half-widths computed with the energy decomposition method are identical or only few 0.1 meV larger with the decomposition method. The differences between widths from the $\textit{gerade}$ and $\textit{ungerade}$ channels are identical with both methods. Notably, there are a few channels where the energy decomposition estimates the partial half-width as up to 1 meV larger. These seem to exclusively affect channels with holes in the inner-valence orbitals 2a$_{1g}$ and 2a$_{2u}$. The contribution from channels involving two outer-valence electrons decreases from 47\%, computed with ACP, to 44\%, computed via energy decomposition.

\begin{figure}
        \includegraphics[width=\linewidth]{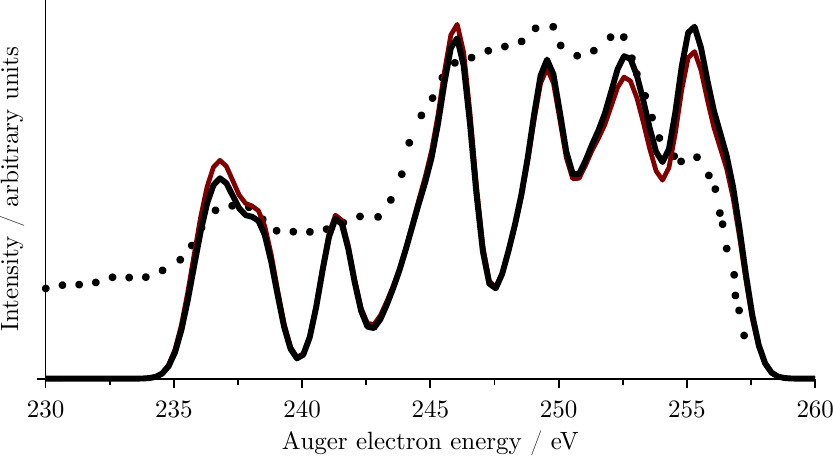}
        \caption{Auger decay spectrum of ethane, computed with ACP-EOMIP-CCSD (black)~\cite{matz23b} and decomposition of the EOMIP energy (red), compared to an experimental spectrum (black dots)~\cite{rye80}. Peaks broadened with a FWHM of 1.5\,eV. Spectra shifted to lower energies by 1.7\,eV to match experiment.}\label{fig:ethan-spektrum}
\end{figure}

Auger spectra can be computed separately for the $\textit{gerade}$ and $\textit{ungerade}$ orbitals, since partial widths and core-ionisation energies result for each hole separately. Figure~\ref{fig:ethan-spektrum} shows the spectra averaged over both states.

The shape of the spectra with the two methods is very similar and follows the experimental spectrum well. The differences can mainly be explained by the energy decomposition predicting larger intensities for lower-energy Auger electrons.

\subsection{Hydrogen sulfide}

Hydrogen sulfide has several core-orbitals, arising from the 1s, 2s, and 2p orbitals of the sulfur atom. The levels have substantially different energies: the 1s ionisation energy is ca. 2500\,eV, while the 2s ionisation energy is only ca. 330\,eV.

Both 1a$_1$ and 2a$_1$ holes can be described with EOMIP-CCSD wave functions as shown in previous work on H$_2$S~\cite{drennhaus24,drennhaus24arxiv}. However, it was shown that different numbers of complex-scaled basis functions are needed to reach basis set convergence. In this work, we used the basis set for which convergence is certain, i.\,e. basis sets with 4 complex-scaled s-, p-, and d-shells for the 1a$_1$-hole, where an optimal $\theta$ of 16$^\circ$ results. Calculations for the 2a$_1$-hole used 8 complex-scaled s-, p-, and d-shells, resulting in an optimal $\theta$ of 14$^\circ$.

For the K-edge hole, we present the widths summed up by the shells of the involved electrons. The absolute and relative numbers for these sums are shown in table~\ref{tab:branchingratios}. The total decay half-width from EOMIP-CCSD calculations amounts to 215.3~meV.~\cite{drennhaus24arxiv}

\begin{table*}
\caption{Branching ratios for the Auger decay of the K-edge hole in hydrogen sulfide, computed with different methods and as expected statistically.}\label{tab:branchingratios}
    \centering
    \begin{tabular}{lrrr|rrrr}\hline
            &\multicolumn{3}{c|}{$\sum\Gamma_{ij}/2$/meV}&\multicolumn{4}{c}{$\sum\Gamma_{ij}/\Gamma$/\%}\\
            Branch&CCSD&\multicolumn{2}{c|}{EOMIP-CCSD}&CCSD&\multicolumn{2}{c}{EOMIP-CCSD}\\
            &decomp.~\cite{drennhaus24arxiv}&ACP~\cite{drennhaus24arxiv}&decomp.&decomp.~\cite{drennhaus24arxiv}&ACP~\cite{drennhaus24arxiv}&decomp.&stat.\\\hline
            L$_1$L$_1$&12.3&10.4&18.4&5.5&5.2&5.5&1.6\\
            L$_1$L$_{2,3}$&62.7&53.0&90.9&28.1&26.5&27.2&9.4\\
            L$_1$M$_1$&2.2&1.8&2.5&1.0&0.9&0.8&3.1\\
            L$_1$M$_{2,3}$&3.8&4.3&5.6&1.7&2.1&1.7&9.4\\
            L$_{2,3}$L$_{2,3}$&124.8&110.0&188.7&55.9&55.1&56.5&14.1\\
            L$_{2,3}$M$_1$&4.1&4.2&6.2&1.8&2.1&1.9&9.4\\
            L$_{2,3}$M$_{2,3}$&12.8&15.3&20.5&5.7&7.7&6.1&28.1\\
            M$_1$M$_1$&0.1&0.1&0.1&0.1&0.0&0.0&1.6\\
            M$_1$M$_{2,3}$&0.4&0.3&0.4&0.1&0.2&0.1&9.4\\
            M$_{2,3}$M$_{2,3}$&0.6&0.5&0.5&0.1&0.2&0.2&14.1\\
        \hline
            LL&199.2&173.5&297.9&89.5&86.8&89.3&25\\
            LM&22.9&25.6&34.8&10.2&12.8&10.4&50\\
            MM&0.6&0.9&1.0&0.3&0.4&0.3&25\\
        \hline
    \end{tabular}
\end{table*}

One can quickly see that while the energy decomposition methods yields noticeably too large partial decay widths for the vast majority of channels, the resulting relative widths are much more similar. Deviations there are mainly larger relative widths of decay channels involving the L-shell orbitals computed with the energy decomposition method. In fact, the relative widths from the EOMIP-CCSD energy decomposition resemble the results from the CCSD energy decomposition more closely than the ACP-EOMIP numbers, but the strong deviation of the sum of partial decay half-width (334 meV) from the total decay half-width (215 meV) make the correctness of any of the absolute partial widths doubtful.

\begin{figure}
        \includegraphics[width=\linewidth]{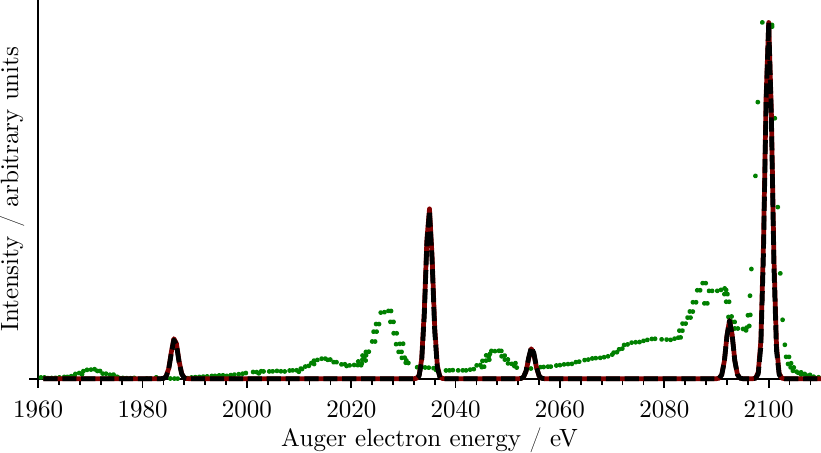}
        \caption{KLL Auger spectrum of the H$_2$S molecule, computed with EOMIP-CCSD and ACP (black dashed line)~\cite{drennhaus24arxiv} and energy decomposition (red line), compared with experimental data~\cite{faegri77} (green dots). Peaks are broadened to 3\,eV FWHM and shifted by 16\,eV to lower energies.}\label{fig:h2skll}
\end{figure}

\begin{figure}
        \includegraphics[width=\linewidth]{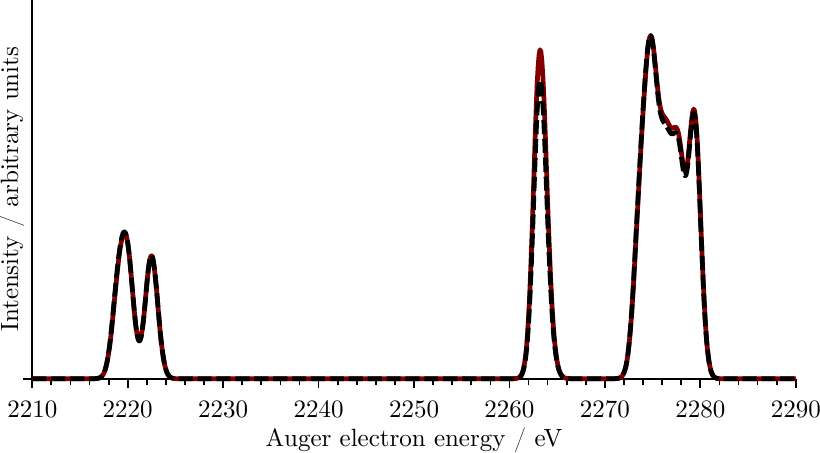}
        \includegraphics[width=\linewidth]{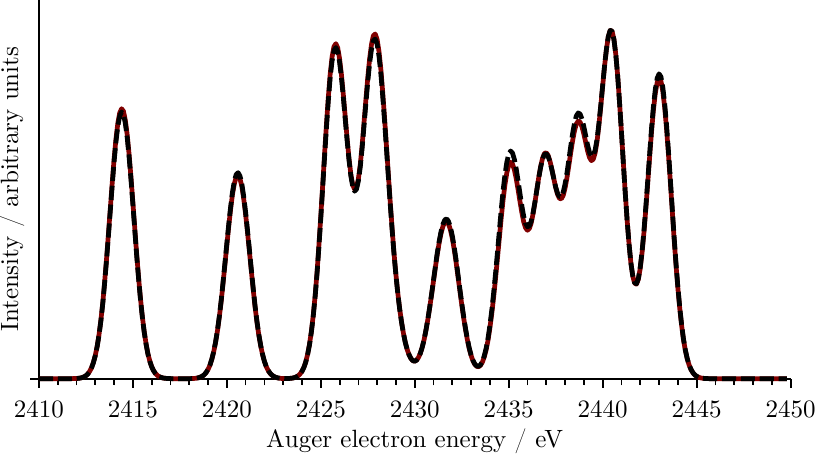}
        \caption{KLM (left) and KMM (right) Auger spectra of the H$_2$S molecule, computed with EOMIP-CCSD and ACP (black dashed line)~\cite{drennhaus24arxiv} and energy decomposition (red line). Peaks are broadened to 1.5\,eV FWHM.}\label{fig:h2skm}
\end{figure}

The different spectral branches are plotted in figure~\ref{fig:h2skll} and~\ref{fig:h2skm}. When normalised to bring the maxima to the same intensity, there is barely any noticeable difference between the ACP and decomposition methods.

The 2a$_1$ core-hole has a much larger decay width due to the possibility of Coster-Kronig decay. This makes up about 97~\% of the decay width of these states.~\cite{drennhaus24} With sufficiently diffuse complex-scaled basis functions, the total half-width of this state with EOMIP-CCSD results to 1672 meV.~\cite{drennhaus24} The partial widths computed for this state with ACP and via energy decomposition are shown in table~\ref{tab:h2s2spw}.

\begin{table}\small
\centering
\caption{Partial Auger decay half-widths in meV for the 2a$_1^{-1}$ state of hydrogen sulfide computed with different methods.}\label{tab:h2s2spw}
\begin{tabular}{lrr|lrr}
\hline
\multicolumn{3}{c|}{L$_{2,3}$M channels} & \multicolumn{3}{c}{MM channels} \\
Decay channel & ACP~\cite{drennhaus24} & decomp. & Decay channel & ACP~\cite{drennhaus24} & decomp.\\
\hline
$^1\textup{B}_2$ (4a$_1^{-1}$1b$_2^{-1}$) & 244.3 & 267.6 & $^1\textup{B}_2$ (4a$_1^{-1}$2b$_2^{-1}$) & 9.3 & 7.4 \\
$^1\textup{B}_1$ (4a$_1^{-1}$1b$_1^{-1}$) & 208.7 & 222.3 & $^1\textup{A}_1$ (4a$_1^{-1}$5a$_1^{-2}$) & 7.5 & 6.7 \\
$^1\textup{A}_1$ (3a$_1^{-1}$4a$_1^{-1}$) & 236.4 & 253.8 & $^1\textup{A}_1$ (4a$_1^{-1}$5a$_1^{-1}$) & 7.2 & 6.3 \\
$^1\textup{A}_1$ (1b$_2^{-1}$2b$_2^{-1}$) & 97.8 & 86.0 & $^3\textup{B}_1$ (4a$_1^{-1}$2b$_1^{-1}$) & 6.9 & 4.7 \\
$^1\textup{A}_1$ (3a$_1^{-1}$5a$_1^{-1}$) & 97.2 & 96.0 & $^3\textup{B}_2$ (4a$_1^{-1}$2b$_2^{-1}$) & 3.8 & 3.5 \\
$^3\textup{A}_1$ (1b$_1^{-1}$2b$_1^{-1}$) & 76.1 & 56.2 & $^3\textup{A}_1$ (4a$_1^{-1}$5a$_1^{-1}$) & 3.3 & 2.9 \\
$^1\textup{A}_1$ (1b$_1^{-1}$2b$_1^{-1}$) & 65.2 & 57.7 & $^1\textup{A}_1$ (5a$_1^{-2}$) & 2.7 & 2.1 \\
$^3\textup{A}_1$ (3a$_1^{-1}$4a$_1^{-1}$) & 57.3 & 42.9 & $^3\textup{B}_1$ (4a$_1^{-1}$2b$_1^{-1}$) & 2.4 & 2.1 \\
$^1\textup{B}_1$ (3a$_1^{-1}$1b$_1^{-1}$) & 59.8 & 44.7 & $^1\textup{B}_2$ (5a$_1^{-1}$2b$_1^{-1}$) & 0.7 & 0.5 \\
$^1\textup{B}_2$ (5a$_1^{-1}$1b$_1^{-1}$) & 43.0 & 49.7 & $^1\textup{B}_1$ (5a$_1^{-1}$2b$_1^{-1}$) & 0.6 & 0.3 \\
$^3\textup{B}_1$ (3a$_1^{-1}$2b$_1^{-1}$) & 34.1 & 25.8 & $^3\textup{B}_2$ (5a$_1^{-1}$2b$_2^{-1}$) & 0.5 & 0.5 \\
$^1\textup{B}_1$ (5a$_1^{-1}$2b$_1^{-1}$) & 33.6 & 37.8 & $^1\textup{A}_1$ (4a$_1^{-1}$2b$_1^{-1}$) & 0.4 & 0.3 \\
$^1\textup{A}_2$ (1b$_1^{-1}$2b$_1^{-1}$) & 32.6 & 22.8 & $^3\textup{B}_2$ (4a$_1^{-1}$2b$_2^{-1}$) & 0.3 & 0.5 \\
$^3\textup{B}_2$ (3a$_1^{-1}$2b$_2^{-1}$) & 32.2 & 22.3 & $^1\textup{A}_1$ (2b$_1^{-2}$) & 0.1 & 0.1 \\
$^3\textup{A}_2$ (2b$_1^{-1}$1b$_2^{-1}$) & 30.6 & 22.6 & $^3\textup{A}_2$ (2b$_1^{-1}$1b$_2^{-1}$) & 0.0 & 0.0 \\
$^3\textup{B}_2$ (5a$_1^{-1}$2b$_1^{-1}$) & 26.7 & 19.4 & $^1\textup{A}_2$ (2b$_1^{-1}$2b$_2^{-1}$) & -0.6 & -0.2 \\
$^3\textup{B}_1$ (5a$_1^{-1}$2b$_2^{-1}$) & 22.9 & 17.6 & & & \\
$^1\textup{B}_1$ (3a$_1^{-1}$2b$_1^{-1}$) & 9.2 & 13.1 & & & \\
$^1\textup{A}_2$ (2b$_1^{-1}$2b$_1^{-1}$) & 8.8 & 10.8 & & & \\
$^1\textup{A}_1$ (4a$_1^{-1}$2b$_2^{-1}$) & 8.3 & 11.9 & & & \\
$^1\textup{B}_2$ (3a$_1^{-1}$1b$_2^{-1}$) & 6.3 & 7.8 & & & \\
$^3\textup{B}_2$ (4a$_1^{-1}$2b$_1^{-1}$) & --5.3 & --0.2 & & & \\
$^1\textup{A}_1$ (3a$_1^{-1}$4a$_1^{-1}$) & --13.7 & --2.9 & & & \\
$^3\textup{B}_1$ (4a$_1^{-1}$1b$_1^{-1}$) & --15.6 & --4.8 & & & \\
Sum&1396.2&1380.9&Sum&45.1&37.7\\\hline
\end{tabular}
\end{table}

The Coster-Kronig (L$_1$L$_{2,3}$M) half-widths computed via energy decomposition show slight deviations of typically 10-20 meV. These are less systematic than before: the most intense channels, decay to singlet states where one hole is in the 4a$_1$, i.\,e. the inner-valence orbital, are overestimated, but the corresponding triplet states are underestimated. The channels for which the ACP method previously yielded negative partial decay widths which was ascribed to remaining basis set incompleteness still possess negative widths, which, however, are calculated much lower. In the sum of the partial widths, the deviations cancel out and the summed Coster-Kronig decay half-width is only 15 meV or 1\% lower with the energy decomposition approach. The non-Coster-Kronig channels are underestimated systematically, but with a larger relative deviation of 16\%. This 2a$_1^{-1}$ state of hydrogen sulfide is the first example where the energy decomposition predicts a smaller sum of partial widths than the ACP method.

\begin{figure}
        \includegraphics[width=\linewidth]{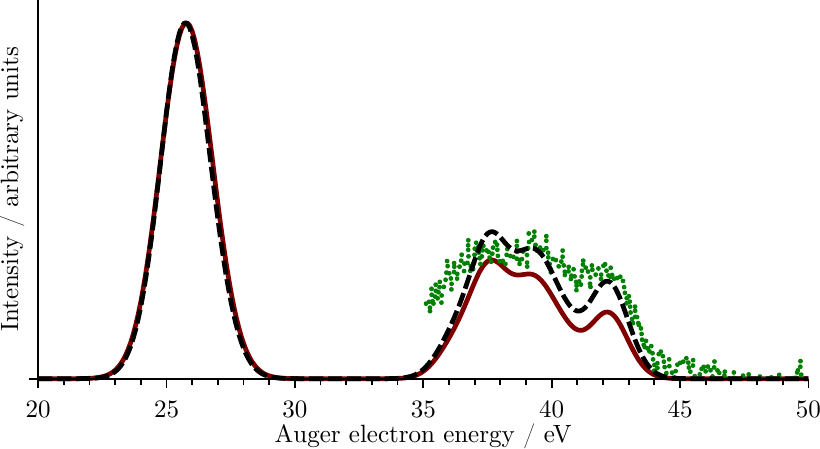}
        \caption{Coster-Kronig spectrum of the H$_2$S molecule, computed with EOMIP-CCSD and ACP (black dashed line)~\cite{drennhaus24} and energy decomposition (red line), compared with experimental data (green dots).~\cite{hikosaka04} Peak broadening with 1.5\,eV, spectrum shifted by 3.1\,eV to higher energies to match experimental data.}\label{fig:h2sck}
\end{figure}

The computed Coster-Kronig spectra are shown in figure~\ref{fig:h2sck} and compared with an experimental measurement of the high-energy signal.

The main difference between the two spectra is the relative intensity of the two signals in the spectrum. The broad structure from 35-45 meV has a lower intensity using the energy decomposition, but otherwise has the same shape that resembles the experimental result.

\begin{figure}
        \includegraphics[width=\linewidth]{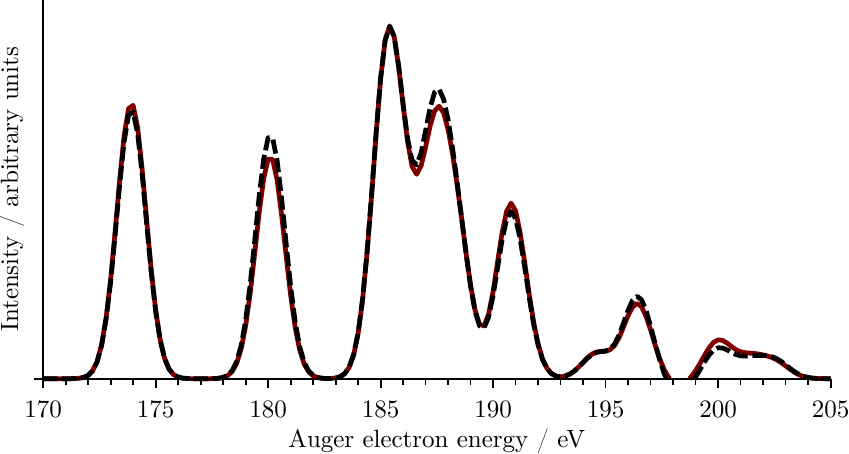}
        \caption{L$_1$MM Auger spectrum of the H$_2$S molecule, computed with EOMIP-CCSD and ACP (black dashed line)~\cite{drennhaus24} and energy decomposition (red line). Peaks are broadened with a FWHM of 1.5\,eV.}\label{fig:h2slmm}
\end{figure}

The L$_1$MM spectrum is shown in figure~\ref{fig:h2slmm}. It has several signals more than the Coster-Kronig spectrum resulting from the larger energetic split of the electronic levels in the M-shell. The differences between the two partial width methods are negligible.

\subsection{The cyanide anion}

Auger decay in the cyanide anion has not been studied using complex-variable methods before. It is also an unprecendented example since it contains different atom types and thus core-holes of very different energies localised on different atoms, and it is an anionic species. The core orbitals at different energy levels result in further open decay channels corresponding to core-core transitions. In the cyanide anion, this so-called interatomic Auger decay involves a $2\upsigma \rightarrow 1\upsigma$ transition.

We determined the optimal scaling angles to 14$^\circ$ for the nitrogen (or 1$\upsigma$) core-hole and 13$^\circ$ for the carbon (or 2$\upsigma$) core-hole. The ionisation energies result as 399.1\,eV (1$\upsigma$) and 285.0\,eV (2$\upsigma$). The total Auger decay half-widths are 37.9\,eV (1$\upsigma$) and 18.2\,eV (2$\upsigma$). The excitation energies match calculated values for core-ionisation processes in similar atoms well, for example 411.3\,eV in dinitrogen~\cite{matz22a}, or 285.0\,eV in methane~\cite{matz23b}, but the total decay widths are much lower than expected -- in previous calculations 62 meV resulted for dinitrogen~\cite{matz22a} and 43 meV for methane~\cite{matz23b}. Whether this is a misestimation of the method or a consequence of the anionic character or electronic structure of cyanide is not obvious due to the lack of an experimental reference.

The partial widths of the cyanide anion are given in table~\ref{tab:cn-pw}.

\begin{table}
\centering
\caption{Partial Auger decay half-widths in meV for the two core-ionised states of the cyanide anion, computed with different methods.}\label{tab:cn-pw}
\begin{tabular}{lcccc}
\hline
        Decay &\multicolumn{2}{c}{N-hole} & \multicolumn{2}{c}{C-hole} \\
channel & ACP & decomp. & ACP & decomp.\\
\hline
        $^3\Sigma$ (2$\upsigma^{-1}$3$\upsigma^{-1}$)&0.001&0.008&\multicolumn{2}{c}{--}\\
        $^3\Sigma$ (2$\upsigma^{-1}$4$\upsigma^{-1}$)&0.024&0.021&\multicolumn{2}{c}{--}\\
        $^3\Sigma$ (2$\upsigma^{-1}$5$\upsigma^{-1}$)&0.018&0.011&\multicolumn{2}{c}{--}\\
        $^3\Pi$ (2$\upsigma^{-1}$1$\uppi^{-1}$)&0.033&0.020&\multicolumn{2}{c}{--}\\
        $^1\Sigma$ (2$\upsigma^{-1}$3$\upsigma^{-1}$)&0.007&0.010&\multicolumn{2}{c}{--}\\
        $^1\Sigma$ (2$\upsigma^{-1}$4$\upsigma^{-1}$)&0.007&0.010&\multicolumn{2}{c}{--}\\
        $^1\Sigma$ (2$\upsigma^{-1}$5$\upsigma^{-1}$)&0.022&0.024&\multicolumn{2}{c}{--}\\
        $^1\Pi$ (2$\upsigma^{-1}$1$\uppi^{-1}$)&0.035&0.024&\multicolumn{2}{c}{--}\\\hline
        $^3\Sigma$ (3$\upsigma^{-1}$4$\upsigma^{-1}$)&1.0&1.2&0.1&0.1\\
        $^3\Sigma$ (3$\upsigma^{-1}$5$\upsigma^{-1}$)&0.3&0.4&0.9&1.2\\
        $^3\Pi$ (3$\upsigma^{-1}$1$\uppi^{-1}$)&1.7&2.0&1.0&1.0\\
        $^1\Sigma$ (3$\upsigma^{-1}$3$\upsigma^{-1}$)&5.0&7.3&2.0&3.0\\
        $^1\Sigma$ (3$\upsigma^{-1}$4$\upsigma^{-1}$)&6.6&9.0&3.0&4.1\\
        $^1\Sigma$ (3$\upsigma^{-1}$5$\upsigma^{-1}$)&2.2&2.8&5.2&6.6\\
        $^1\Pi$ (3$\upsigma^{-1}$1$\uppi^{-1}$)&8.8&10.5&5.4&5.9\\
        $^3\Sigma$ (4$\upsigma^{-1}$5$\upsigma^{-1}$)&0.1&0.2&0.4&0.5\\
        $^3\Pi$ (4$\upsigma^{-1}$1$\uppi^{-1}$)&0.5&0.6&0.5&0.6\\
        $^1\Sigma$ (4$\upsigma^{-1}$4$\upsigma^{-1}$)&3.0&3.9&1.2&1.5\\
        $^1\Sigma$ (4$\upsigma^{-1}$5$\upsigma^{-1}$)&2.4&3.0&2.7&3.3\\
        $^1\Pi$ (4$\upsigma^{-1}$1$\uppi^{-1}$)&8.4&9.5&2.2&2.4\\
        $^3\Pi$ (5$\upsigma^{-1}$1$\uppi^{-1}$)&0.0&0.0&0.2&0.2\\
        $^1\Sigma$ (5$\upsigma^{-1}$5$\upsigma^{-1}$)&0.6&0.7&3.6&4.3\\
        $^1\Pi$ (5$\upsigma^{-1}$1$\uppi^{-1}$)&3.8&4.1&7.2&7.1\\
        $^3\Delta$ (1$\uppi^{-2}$)&0.0&0.0&0.0&0.0\\
        $^1\Sigma$ (1$\uppi^{-2}$)&2.3&2.3&1.0&1.0\\
        $^1\Delta$ (1$\uppi^{-2}$)&8.9&8.7&4.5&4.0\\\hline
        Sum&55.7&66.3&41.0&46.7\\\hline
\end{tabular}
\end{table}

Despite the total decay widths being lower than expected, the partial widths add up to numbers much closer to the expectation both with ACP and with energy decomposition. Similar to the numbers in the hydrocarbons, the ACP widths are lower than these typical total decay widths and the numbers obtained via energy decomposition are larger and the deviations are mainly explained by energy decomposition computing up to 2\,meV larger decay half-widths for channels that involve the inner-valence orbitals 3$\upsigma$ and 4$\upsigma$. However, the very different total decay widths for cyanide imply that there are substantial contributions from determinants that do not describe Auger decay.

The interatomic Auger decay channels are all of negligible width, adding up to just 0.15 meV, computed with ACP, and 0.13 meV, computed with energy decomposition, corresponding to 0.2-0.3\% of the total decay width. This order of magnitude has been measured before for interatomic Auger decay in carbon monoxide.~\cite{ogurtsov07}

The vast majority of Auger decay happens from the valence orbitals, where very different partial widths are observed for the two initial core-hole states: While the sum of total widths is larger for the nitrogen hole, this is not a systematic trend over all widths. In fact, partial widths for most channels are much larger in one core-ionised state than in the other.

The criterion for if decay via a channel is more or less probable in the carbon hole is the involvement of the 5$\upsigma$ orbital, the highest-energy occupied orbital. Channels that include it make up 17\% of the decay width of the nitrogen core-hole, but 49\% for the carbon core-hole. This orbital is centered mainly at the carbon atom and plays an important role in the binding of cyanide as a ligand. The localisation of the orbital can give a partial explanation for why it is more involved in the decay of the core-hole localised on carbon.

\begin{figure}
        \includegraphics[width=\linewidth]{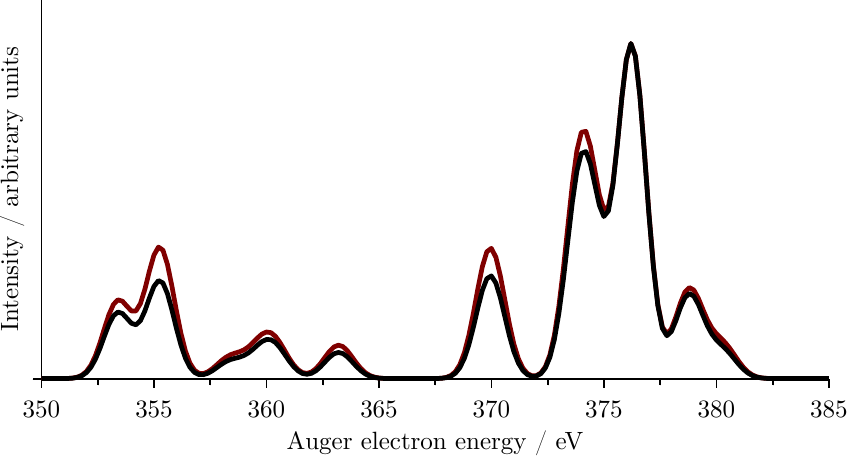}
        \caption{Auger decay spectrum of the 1$\upsigma$ core-hole (centered on the nitrogen atom) of the cyanide anion, computed with EOMIP-CCSD and ACP (black) and energy decomposition (red line). Peaks are broadened with a FWHM of 1.5\,eV.}\label{fig:cn-n-spek}
\end{figure}

\begin{figure}
        \includegraphics[width=\linewidth]{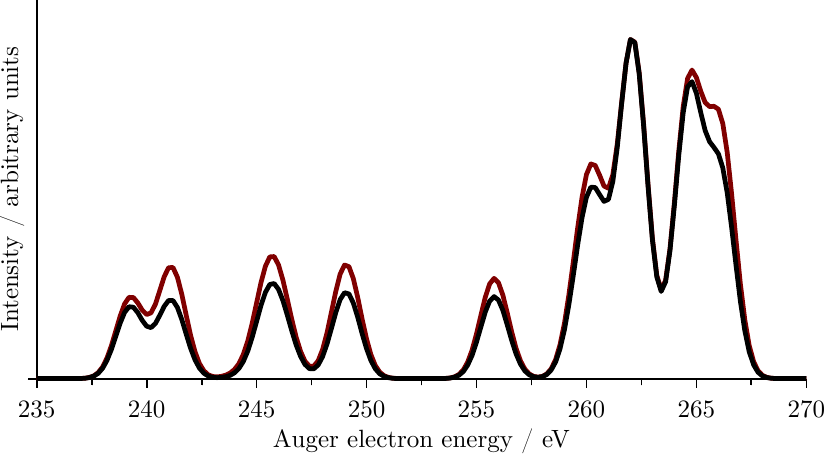}
        \caption{Auger decay spectrum of the 2$\upsigma$ core-hole (centered on the carbon atom) of the cyanide anion, computed with EOMIP-CCSD and ACP (black) and energy decomposition (red line). Peaks are broadened with a FWHM of 1.5\,eV.}\label{fig:cn-c-spek}
\end{figure}

Double ionisation energies to describe the final states were computed in order to generate spectra, which are shown for the two core-holes in figures~\ref{fig:cn-n-spek} and~\ref{fig:cn-c-spek}.

As can be expected from the very different set of partial decay widths, the spectra of the two core holes are very dissimilar. The spectrum arising from the nitrogen core-hole is dominated by a large double peak around 375\,eV, mainly produced by double ionisation from the $\uppi$ orbitals. A signal of lower intensity around 370\,eV stems from the 4$\sigma^{-2}$ decay channel, while inner-valence processes lead to shallow signals between 353 and 364\,eV.

At the high-energy end of the nitrogen-edge spectrum, around 379\,eV, is a quite low-intensity signal, corresponding to ionisation of the HOMO. The corresponding peak in the carbon-edge spectrum, at ca. 265\,eV, is much more intense and of similar intensity as the signal from the $\uppi$ orbitals around 260-263\,eV, a direct signature of the large decay widths off channels involving the HOMO. The lower-energy part of the spectrum has a higher intensity relative to the higher-energy part, but none of the signal dominates significantly. In line with the numerical results, it is also noticeable how the curve from the density matrix decomposition predicts a larger intensity for the low-energy end of the spectrum.

\begin{figure}
        \includegraphics[width=\linewidth]{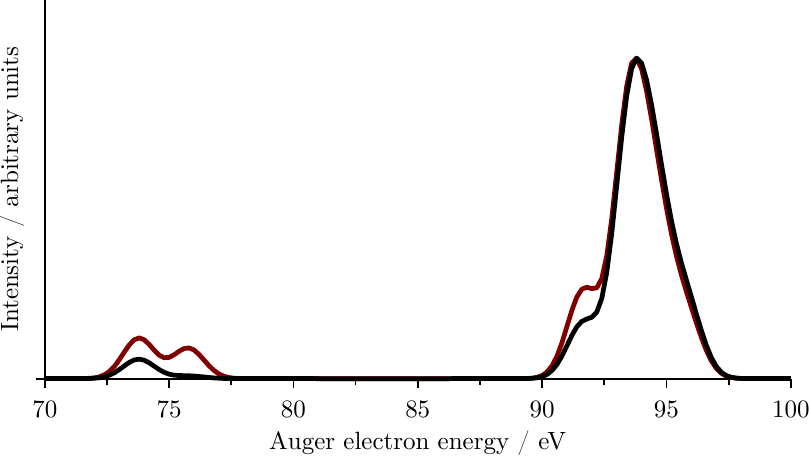}
        \caption{Interatomic Auger decay spectrum for the 1$\upsigma$ core-hole in the cyanide anion, computed with EOMIP-CCSD and ACP (black) and energy decomposition (red line). Peaks are broadened with a FWHM of 1.5\,eV.}\label{fig:cn-interatomic}
\end{figure}

Figure~\ref{fig:cn-interatomic} shows the interatomic Auger decay spectrum. It contains only two signals with very different intensities: the high-energy signal from 91 to 97\,eV with a shoulder around 92\,eV is dominant, while the low-energy signal around 75\,eV has only a fraction of its intensity, where energy decomposition predicts two shallow maxima and ACP only one. In previous calculations and experiments~\cite{ogurtsov07}, similar intensities had been found for inner- and outer-valence interatomic Auger processes, and it should be noted that the partial widths of these signals are outside the typical accuracy of the EOMIP-CCSD method and an intensity distribution obtained from them should be viewed critically.

\section{Conclusion}

In this manuscript, we presented a novel approach for the computation of partial widths for EOMIP-CCSD wave functions in combination with complex-scaled basis functions, evaluating them only from the density matrix. This reduces the computational cost of obtaining partial decay widths because ACP calculations, which have to be repeated for each channel, can be avoided. These computational savings are moderate for molecules with few decay channels or those where EOMIP-CCSD calculations with the Davidson diagonalisation algorithm converge within few or few tens of iterations. However, for some electronic states even of small systems, this convergence is much harder to establish, for example for the L$_\mathrm{1}$-edge ionised state of H$_2$S, which requires 130 iterations for the EOMIP-CCSD calculation in the full excitation manifold and typically 500-700 for each ACP-EOMIP-CCSD calculation.

The method is also of conceptual interest because it allows a direct extraction of the contribution of specific electronic transitions from the wave function via computing the density matrix. This yields decay widths that are additive, unlike the ACP approach, where the remaining wave function can adapt to the missing channel, introducing errors to the computed partial decay widths. Typically, this leads to underestimated partial widths since the other channels become more intense to ``fill in'' for the missing channel. This effect seems to be stronger for channels involving deeper-lying orbitals, where the relative widths computed with the energy decomposition are larger than the ACP widths in most cases. Ultimately, it is not clear if this is rather a shortcoming of the ACP method or of the energy decomposition approach.

Besides this, the spectra computed from the density matrix are almost indistinguishable from those computed with the ACP for all molecules and states studied here. We note that in the K-edge ionised hydrogen sulfide, the sum of partial widths from energy decomposition is in disagreement with both the total decay width and the ACP result, but the spectra look identical upon normalisation of the intensities.

We conclude that the method is highly promising to evaluate Auger decay spectra in all core-holes that can be described with EOMIP-CCSD wave functions. It might also be a valid approach for different types of electronic resonances: inner-valence ionised clusters which can decay via intermolecular Coulombic decay, for which no methods for the computation of partial widths are currently available~\cite{parravicini23}, or superexcited states that are described with EOMEE-CCSD wave functions, including core-excited states undergoing resonant Auger decay~\cite{armen00}. This method can also be used to gain more understanding about the effects that Auger Channel Projectors have on the decay widths by providing a second point of reference to evaluate contributions to the total decay widths.

We hope that this method will enable more investigations of the partial widths of electronic resonances, which give useful information about the state of the system after the electronic decay. Since density matrices are quantities of general interest in computational chemistry, this method is potentially also extendable to other wave functions approaches. Unfortunately, its application to decay processes involving more than two electrons is not trivial, since these transitions do not directly contribute to the correlation energy, but also have no unambiguous entries corresponding to them in the two-electron density matrix. To compute the widths of such transitions, ACP-like methods might still remain the preferrable option.

Finally, we also applied the CBF-EOMIP-CCSD approach to the cyanide anion, which is a novel type of system in several ways. Total widths resulting from these calculations are unexpectedly low for the atom types involved, but the partial widths sum up to numbers comparable to the widths of core-holes in carbon and nitrogen that have been measured and computed for other molecules. Both methods reproduce the very different distribution of the total width over the decay channels depending on the initial core-hole: the core orbital of the carbon atom tends to decay involving the carbon-centered HOMO, leading to more intensity at the high-energy end of the spectrum, while such channels are improbable for decay of the nitrogen core hole. We also investigate the possibility of interatomic Auger decay, where the carbon core electrons refill core holes localised at the nitrogen atom, but found that this decay process only accounts for 0.2-0.3\% of the total decay width.

The observed selectivity for the holes in the final states might have interesting implications for the electronic structure of complexes involving cyanide as a ligand, since the different core ionisation processes might leave the system in a chemically different state. This warrants further investigation of Auger decay in other ligand molecules or of entire complexes, which can be carried out with less resources required with the newly established energy decomposition method for EOMIP-CCSD wave functions.

\section*{Acknowledgments}
The authors are thankful for Prof. Sonia Coriani for helpful discussions and comments regarding the manuscript.

T.-C. J. gratefully acknowledges funding from the European
Research Council (ERC) under the European Union’s Horizon
2020 research and innovation program (Grant Agreement No.
851766) and the KU Leuven internal funds (Grant No. C14/22/083).

\section*{References}
\vspace{-1em}
\bibliographystyle{ieeetr}
\bibliography{aipsamp}
\newpage\clearpage
\section*{Supporting Information}

\begin{table}[h]\centering
\caption{Used exponents for complex basis functions by atom type and angular momentum.}\label{tab:appendix}
\begin{tabular}{llll}\hline
Atom&s&p&d\\\hline
\multirow{8}{*}{H}&0.330271317&1.329389259&1.808464179\\
&0.1069882&0.431001171&0.319694337\\
&0.2139764&0.862002265&0.639388599\\
&0.0534941&0.215500547&0.159847169\\
&0.02674705&0.107750274&0.079923584\\
&0.013373525&0.053875137&0.039961792\\
&0.006686762&0.026937568&0.019980896\\
&0.003343381&0.013468784&0.009990448\\\hline
\multirow{3}{*}{C}&1.5775008&6.349666&8.637909\\
&0.5110162&2.0586245&1.526981\\
&0.2555081&1.0293123&0.7634905\\\hline
\multirow{3}{*}{N}&2.1804898&8.7767831&11.9396915\\
&0.7063487&2.8455199&2.1106592\\
&0.3531745&1.4227600&1.0555330\\\hline
\multirow{8}{*}{S}&2.381349933&9.585273863&13.03954\\
&0.771415287&3.107640767&2.305086907\\
&1.542830574&6.215280984&4.610173265\\
&0.385707644&1.553820109&1.152543454\\
&0.192853822&0.776910054&0.576271727\\
&0.096426911&0.388455027&0.288135863\\
&0.048213455&0.194227514&0.144067932\\
&0.024106728&0.097113757&0.072033966\\\hline
\end{tabular}
\end{table}

\end{document}